# Prediction of Flavin Mononucleotide (FMN) Binding Sites in Proteins Using the 3D Search Motif Method and Double-Centroid Reduced Representation of Protein 3D Structures*

**Arkanjan Banerjee[1] and Vicente M. Reyes, Ph.D.[2], [3]**

(*, M. S. Thesis; [1], M. S. student; [2], thesis advisor)
[3], E-mail: **vmrsbi.RIT.biology@gmail.com**

**Arkanjan Banerjee**

Dept. of Biological Sciences, School of Life Sciences
Rochester Institute of Technology
One Lomb Memorial Drive, Rochester, NY 14623

April 2012

# ACKNOWLEDGEMENTS

I would like to express my gratitude for all those who gave me the possibility to complete the thesis. My time here in RIT was a time where I got to learn a lot in a diverse environment. From having a basic understand of the subject I went on to get a clear understanding of the current advances in the field. Not only did I pick up technologies and skills I had hoped to master in the past but I also got the opportunity to work with some bright minds and amazing people.

I am especially thankful to the Department of Biological sciences (Bioinformatics) for teaching me most of the skills I have today and also encouraging me to be creative. I learnt a lot from my professors and a huge part of what I am is because of them.

I would like to express my most sincere gratitude to my thesis advisor Dr. Vicente Reyes who was with me through the ups and downs of my work. He was a mentor that gave me the strength to get keep going when things had gone terribly wrong. I exchanged hundreds of emails and he was always the one to reply instantly and guide me when I had doubts. I learnt a lot from him and this thesis is what it is because of him. I would also like to express my gratitude to Dr. Gary Skuse and Dr. Paul Craig for giving me advice on how to proceed with my thesis and evaluating my work.

In the course of my thesis I faced something which very few students have faced. A hardware failure had caused a loss of all my data at one point and I would like to thank Dr. Michael Osier and Kyle Dewey for helping me recover that data. They worked really hard to get my work back to where it was as quick as possible.

I would also like to thank Dr. Gurcharan Khurana, Ryan Lewis and Thomas Batzold and the other members of RIT Research Colloboratory for allowing me to use their advanced computational resources.

Lastly I would like to thank my parents and all my friends for supporting me and encouraging me and once again thank you RIT for letting me be part of the RIT family.

# Abstract

A pharmacophore consists of the parts of the structure of the ligand that are sufficient to express the biological and pharmacological effects of the ligand. It is usually a substructure of the entire structure of the ligand. Small organic molecules called ligands or metabolites in the cell form complexes with biomolecules (usually proteins) to serve different purposes. The sites at which
the ligands bind are known as ligand binding sites, which are essentially "pockets"
which have complementary shapes and patterns of charge distribution with the ligands. Sometimes a pocket is induced by the ligand itself. If we study different bound conformations of ligands it is found that they share a specific 3 dimensional pattern that is more or less common and is responsible for its binding and which is complimentary in 3 dimensional geometry and charge distribution pattern with its cognate binding site in the protein.

This work studies the three dimensional structure of the consensus ligand binding site for the ligand FMN. A training set for the ligand binding sites was made and
a 3D consensus binding site motif was determined for FMN. The FMN system was studied and its binding sites in its respective regulator proteins. The ability to identify ligand binding site by scanning the 3D binding site consensus motif in protein 3D structures is an important step in drug target discovery. Once a pharmacophore template is found it can also be used to design other potential molecules that can bind to it and thus serve as novel drugs.

**Tables**:

# INTRODUCTION

The pharmacophore for a given ligand includes the parts of the structure of the ligand that are sufficient to express the biological and pharmacological effects of the ligand. Usually only a substructure of the whole structure of the ligand molecule is responsible for the biological effects of a ligand. When a ligand binds to a protein molecule it binds to a specific 3 dimensional conformation of the ligand binding site in the protein which is known to be conserved for each and every ligand[8].The ligand binding site essentially serves as a pocket that has a complementary shape and charge distribuition with the ligand.  When a protein molecule is unfolded the sites responsible for the ligand binding can be present in different parts of the protein molecule. However on protein folding the sites form the specific 3 dimensional pocket that is complementary in shape and charge distribution to the ligand molecule that is binding to the protein and forms the ligand binding site for that particular ligand. (Fig. 1)

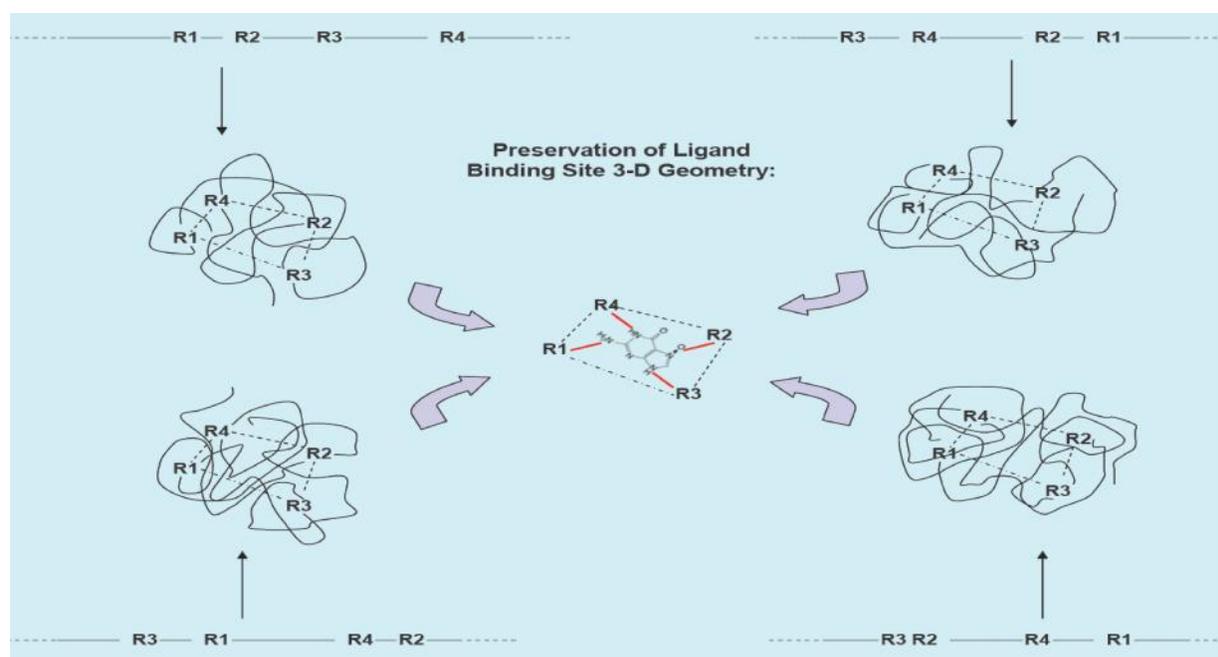

**Figure 1: Preservation of Ligand binding site 3-D Geometry.** The figure shows a protein that has different secondary structures. Unfolded they have the four sites responsible for binding the ligand namely R1, R2, R3 and R4 in different sections of the protein. However on folding they form the tetrahedral model as shown in the figure [11].

Several research groups, including N Schormann et al. [17] have focused on a structure based approach to identify pharmacophores of a particular ligand. Studying the 3 dimensional structure of the ligand binding site of receptor protein molecules for a particular ligand can lead to pharmacophore identification for a particular ligand. Pharmacophore identification can subsequently lead to the discovery of novel molecules that can mimic the ligand and lead to drug discovery. For example a simple illustration of binding site will be [3]:

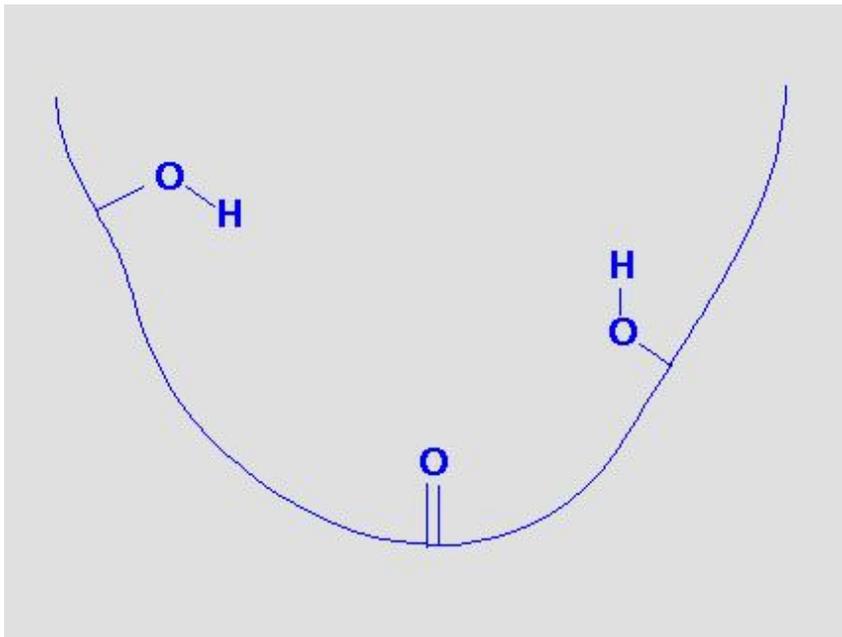

Figure 2:Ligand Binding Site[4]

There are obvious sites for hydrogen bonds and van der Waal's interactions and it can be postulated that a molecule that could bind to this site would provide functional groups that could hydrogen bond to the receptor functional groups. The ligand molecule and its corresponding binding site in the protein must be complementary in both 3D geometry (architecture) and charge distribution pattern.

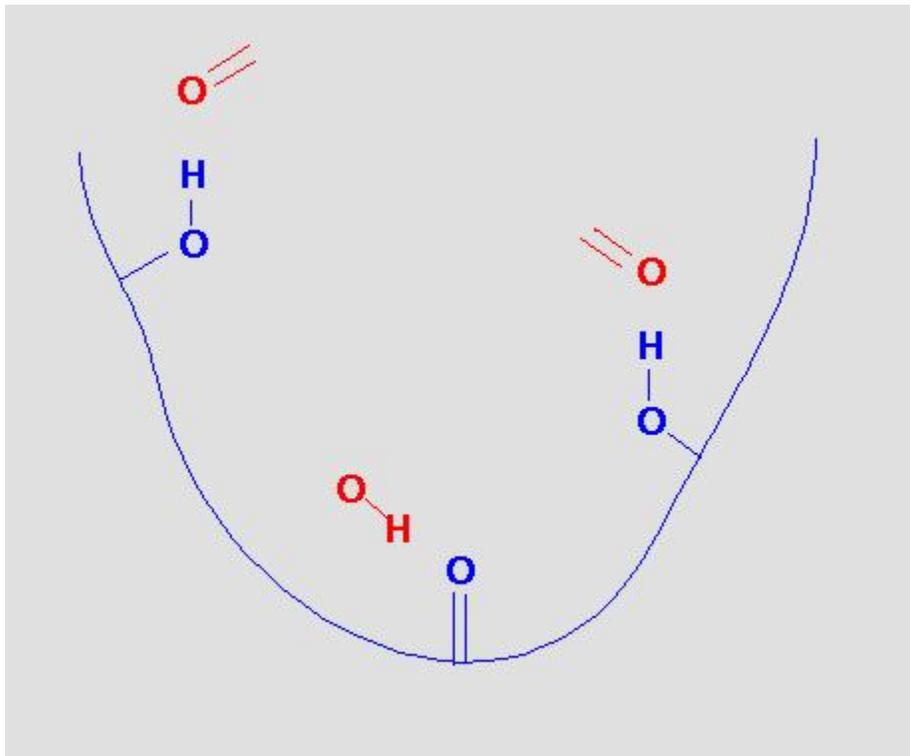

**Figure 3:Ligands binding to the ligand binding site[4]**

From identification of the ligand binding sites (the blue structure), novel molecules can be designed that have functional groups that are complimentary to the binding site residue like:

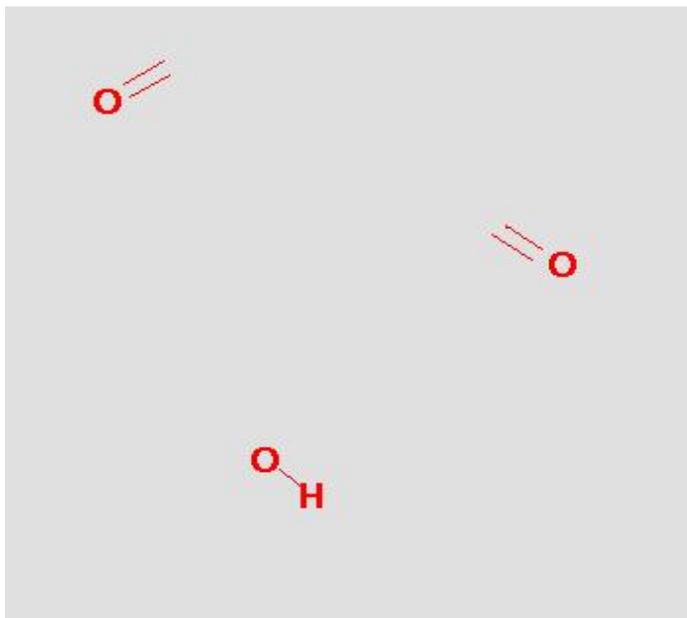

**Figure 4: Pharmacophore[4]**

Keeping this in mind different molecules that can bind to the particular pharmacophore can be searched for. The search may give molecules like:

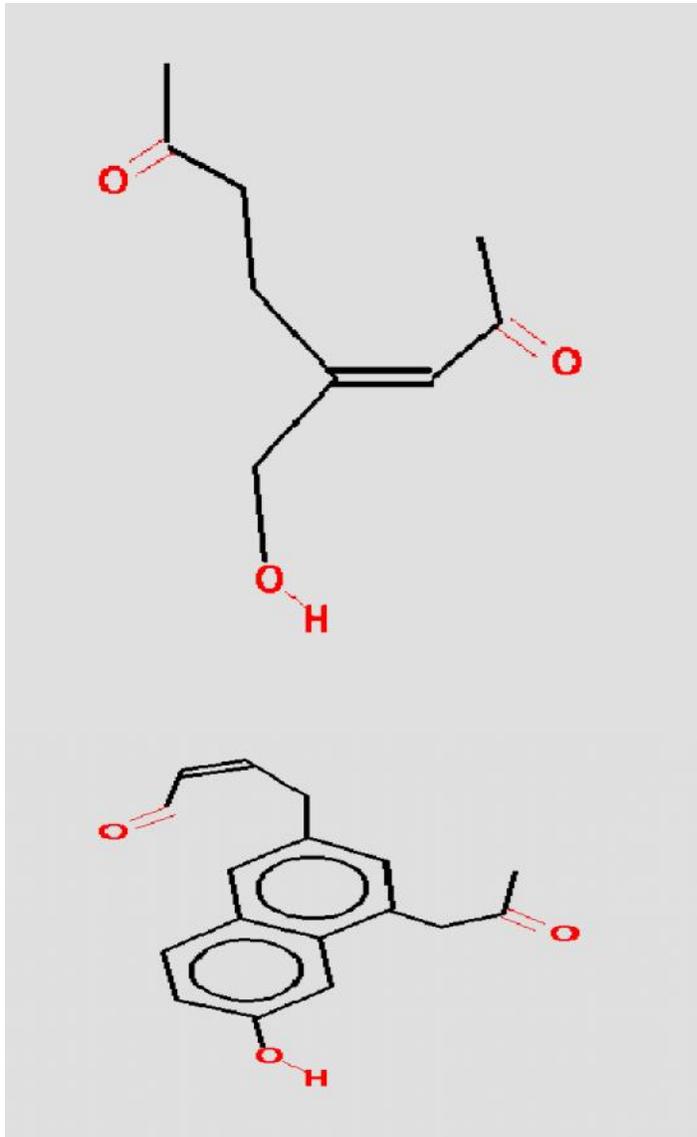

**Figure 5:Possible molecules binding to the ligand binding site[4]**

For our research we seek to identify the specific 3 dimensional structure responsible for the binding of the ligand FMN by studying proteins that have FMN bound to them. The DCRR or double centroid reduced representation [11] is used to study the protein structures to identify a common 3 dimensional motif. The reason DCRR was used is because the default AAR or all atom representation gives too much information and makes it very difficult and complicated to correctly identify a 3D motif. DCRR on the other hand minimizes the information presented while retaining the amount of information needed to identify the conserved 3 dimensional motif.

It was found by Dr. Vicente Reyes and Vrunda Seth in their research [11] that for small ligands like GTP, ATP and FMN the 3 dimensional motif in the ligand binding site can be accurately represented as a tetrahedral structure with the four most dominant amino acids in the side chain or backbone interacting with the ligand, making up the four edges of the tetrahedron. If the 3 dimensional motif was represented as a triangular structure it could not accurately capture a lot of information and led to a lot of false positives while screening for ligand binding sites. However anything more than a tetrahedral model was just unnecessarily complicating the 3 dimensional motif identification process while giving the same results as when a tetrahedral structure was used. Thus for identifying the fixed 3 dimensional ligand binding site responsible for binding FMN we study some structures that have FMN bound to it and try and construct a tetrahedral motif that is common to all the proteins. The proteins that are studied that are known to have FMN bound to it make up what is known as the training set. Fig.6 illustrates few protein structures that contain a bound ligand and identify the fixed 3 dimensional motif responsible for ligand binding by representing it as a tetrahedron.

| | | | |
|---|---|---|---|
| ligand bound to protein receptor 1 | 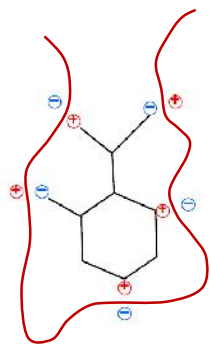 | ligand bound to protein receptor 2 | 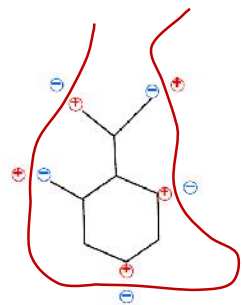 |
| ligand bound to protein receptor 3 | 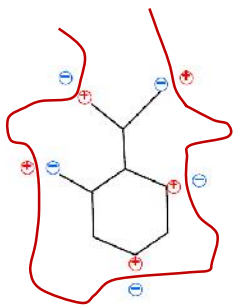 | ligand bound to protein receptor 4 | 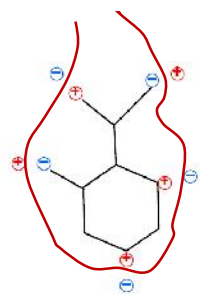 |

**Figure 6: Training set for finding pharmacophores [4]**

For example based on the four training structures on the previous page, the consensus binding site for the ligand constructed in Fig. 7:

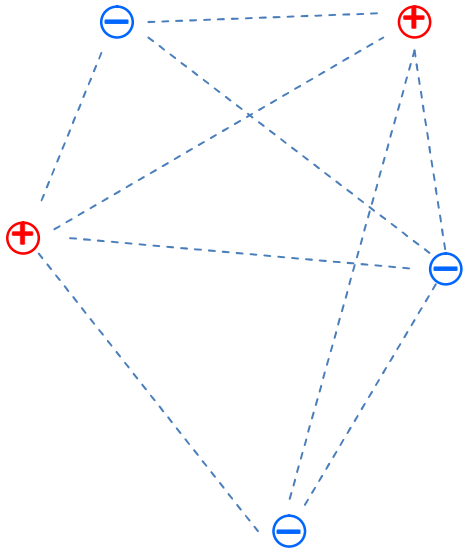

**Figure 7: Mathematical model for the ligand binding site [4]**

Once the fixed 3 dimensional motif was found by studying the training structures an algorithm written in FORTRAN was utilized to identify if that 3 dimensional motif was present in protein structures passed into the algorithm. The algorithm was first trained and tested on training structures having known ligand interactions and then subsequently used on unannotated structures in PDB with unknown function to identify possible candidates that might have FMN bound to them.

Given below is a description of FMN. The functional groups marked in blue in the structure are good candidates as contact points of the receptor protein and was used to construct the consensus binding site of the pharmacophore.

1. FMN:

    FMN or flavin mononucleotide functions as a prosthetic group for several oxidoreductases. It is a stronger oxidizing agent than NAD and is derived from riboflavin like FAD [20].

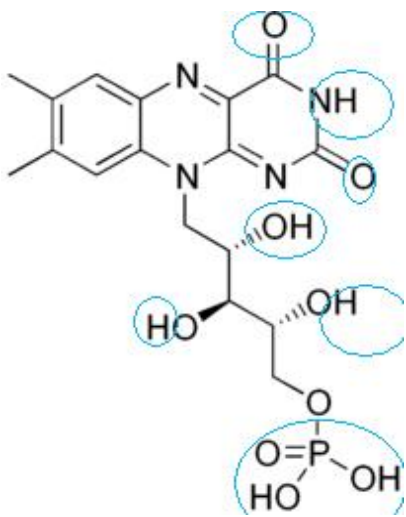

Figure 8: FMN

# MOTIVATION:

Our aim is to determine the 3 dimensional consensus binding site motif which will help in the design of novel molecules that will bind to those sites and thus serve as possible novel drugs. Drugs can either act as agonists or antagonists for the particular binding site.

**Agonists** are chemicals that bind to the receptor and trigger a response by the cell. Sometimes the agonist can trigger a response that is more than the original ligand would have produced. These agonists are called as **super- agonists**. At other times the agonist may do the opposite and trigger a lower response. These agonists are called **inverse-agonists**. A **receptor antagonist** is basically an inhibitor that binds to the receptor site thus preventing the ligand from binding to it and triggering a response.

Though pharmacophore studies have been conducted before, most of them are based on the all atom representation for proteins. In our study however this work uses the double centroid reduced representation [2]. Using the DCRR representation makes the analysis of the 3 dimensional structures easier because of the display presented by DCRR. The AAR model display can be overwhelming and that makes it difficult to model a motif from it. However DCRR captures just enough chemical information to accurately display the secondary structure of a protein.

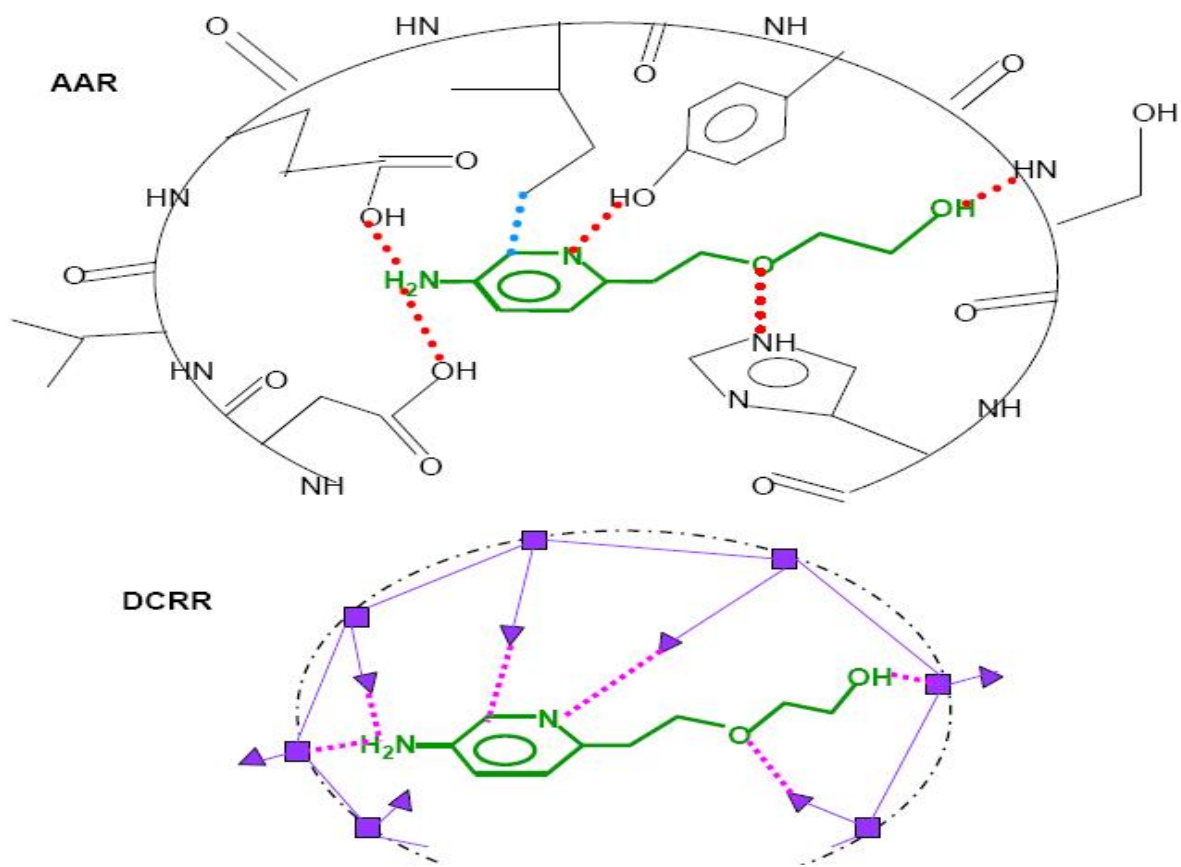

**Figure 9: Finding pharmacophores in DCRR as opposed to AAR.** The DCRR reduces the information presented as shown in the figure and makes it easier to construct a 3 dimensional motif of the ligand binding sites [11].

## SPECIFIC AIMS:

*Specific Aim #1.* Using a Training Set of a dozen solved structures with bound FMN, Determine and Analyze the Characteristics of the Binding Site for FMN and build a 3D Binding Site tetrahedral Consensus Motif for use in Screening.

*Specific Aim #2.* Screen a Dataset Composed of Most of the Current Functionally Unannotated Structures from the PDB for Screening using the screening procedure established for the above mathematical model for the binding site.[11]

## METHODS AND RESULTS:

The whole process can be subdivided into three broad categories

a) Gather data that can be used to prepare the screening algorithm

b) Train the screening algorithm by testing it on a predefined training set

c) Use the algorithm to screen functionally unannotated structures in PDB

## Gather data to prepare the screening algorithm

The following flowchart elucidates the steps that were used to gather this data

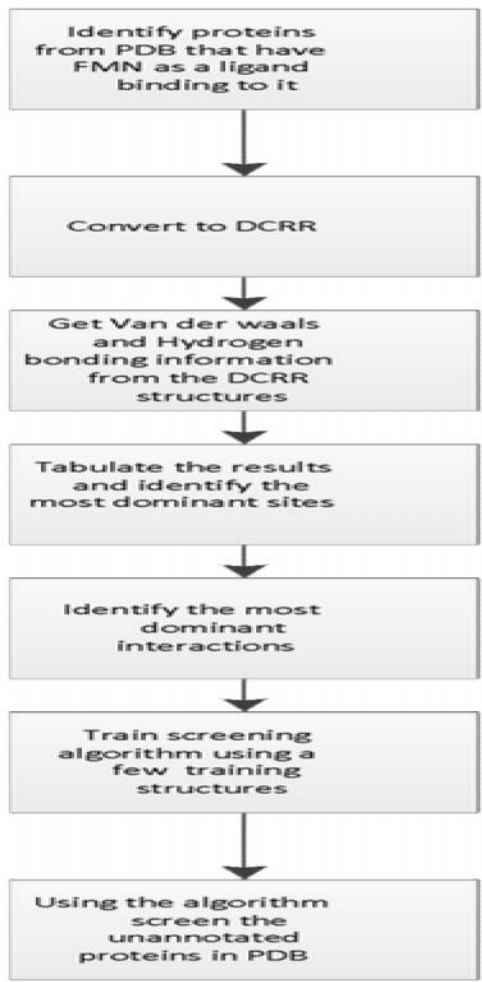

**Figure 10: Flow of the whole process:** The proteins to be used as the training structures are first identified and converted to DCRR. After that the Van der Waals and Hydrogen bonding information is obtained by running FORTRAN programs that determine this information. From this the most dominant atoms and interactions are identified and used as qualitative and quantitative inputs for the screening algorithm. Once this screening algorithm is ready it is trained by testing it on structures known to have FMN bound to them. The last step of the process is to use the screening algorithm to screen unannotated structures in PDB

**Identifying proteins from PDB that have FMN as a ligand bound to it:**

The PDB database was searched to find sixteen proteins that actually have FMN bound to it. These proteins were used to study the structures of the proteins and then determine the tetrahedral motif that was highlighted as being common to these proteins. The following proteins were selected

| FMN | |
|---|---|
| **PDB ID** | **MOLECULE NAME** |
| 1D3G | DIHYDROOROTATE DEHYDROGENASE |
| 1B1C | PROTEIN (NADPH-CYTOCHROME P450 REDUCTASE) |
| 1NRG | pyridoxine 5'-phosphate oxidase |
| 2BXV | DIHYDROOROTATE DEHYDROGENASE |
| 2RDU | Hydroxyacid oxidase 1 |
| 2RDW | Hydroxyacid oxidase 1 |
| 2PRH | Dihydroorotate dehydrogenase, mitochondria |
| 2RDT | Hydroxyacid oxidase 1 |
| 3HR4 | Nitric oxide synthase, inducible |
| 3KVJ | Dihydroorotate dehydrogenase, mitochondria |
| 2NZL | Hydroxyacid oxidase 1 |
| 2FPY | Dihydroorotate dehydrogenase, mitochondrial precursor |
| 1QZU | hypothetical protein MDS018 |
| 3KVK | Dihydroorotate dehydrogenase, mitochondrial |
| 3KVM | Dihydroorotate dehydrogenase, mitochondrial |
| 2WKP | NPH1-1, RAS-RELATED C3 BOTULINUM TOXIN SUBSTRATE 1 |

Table 1: Training set

All the selected proteins are proteins that are found in humans so that we can eliminate any irregularity arising from diverse species.

**Converting to DCRR**

Since this work utilizes the benefits of using a DCRR structure as opposed to a AAR structure as discussed above one of the very important steps is to convert the structures into DCRR ones from the default AAR structure given in PDB. There can be two ways of doing this. The first way is to use the DCRR website and give the protein name and convert it from there. The second way is to use the programs and scripts written by Vrunda Seth and Dr. Vicente Reyes for their work involving conversion of AAR structures into DCRR structures.
For this thesis the second approach was used this is primarily because of three reasons.

1. The DCRR website is not comprehensive. It does not have all the structures that are present in PDB. Most of the proteins used in this work was not present in the website.

2. A lot of additional files generated by the DCRR conversion program were used in this work. This is primarily related to the Van der Waals and hydrogen bonding information in the proteins. These text files are not available in the website.

3. Getting the information from a Unix shell and a fast processor was faster than relying on the internet.

**Get Van der waals and Hydrogen bonding information from the DCRR structures**

Running the DCRR algorithm in the above steps gave the Van der Waals and hydrogen bonding information in separate files. These files were combined to get the ligand interactions for each and every protein. Basically to find out the Hydrogen bonding information a nearest neighbor analysis was used to identify H-bonding and VDW interactions. A sphere typically of radius 5.0 - 6.0 Å was constructed around every atom in the protein as center; all other atoms found in the interior of such a sphere is considered 'neighbors' of the central atom. Hydrogen bonds were taken to be those that are within close neighborhood of 2.80 Å between central atom and neighbor, with the compatible chemical identities (those involving P, O, N and/or S). As for van der Waals interaction, we considered only those of the C-H• • • • H-C type and whose distances between the carbon atoms are within close neighborhood of 3.38 Å[11][21]

After this was done the distance of the ligand molecule to the amino acid backbone or side chain was calculated using the distance formula $\sqrt{(x-x_0)^2 + (y-y_0)^2 + (z-z_0)^2}$ applied to the Cartesian coordinates of the amino acid in the protein and the ligand molecule. See Fig. 11

**Tabulate the results and identify the most dominant binding sites**

Once the ligand interactions were extracted from the different proteins a spreadsheet was made to carefully study the proteins and identify a tetrahedral model to be used for screening

|   | 1B1C | 1D3G | 1NRG | 1QZU | 2BXV | 2FPY | 2NZL | 2PRH | 2RDT | 2RDU | 2RDU | 2WKP | 3HR4 | 3KVJ | 3KVK | 3KVM |
|---|---|---|---|---|---|---|---|---|---|---|---|---|---|---|---|---|
|   |   |   |   |   | **TRAINING STRUCTURES** | | | | | | | | | | | |
| C5' | ALA-599(s) 3.74 | LEU-2113(s) 5.02 VAL-2308(s) 4.88 GLY- |   |   | LEU-2098(s) 5.09 GLY-2291(b) 4.44 | LEU-2128(s) 5.10 GLY-2321(b) 4.37 |   | LEU-2070(s) 4.97 | ASP-2061(s) 3.62 | ASP-2272(s) 3.66 | GLY-2284(b) 4.57 | VAL-497(s) 4.88 ARG-536(s) 5.30 | LYS-8107(s) 4.19 | LEU-2122(s) 5.00 GLY-2318(b) 4.47 | LEU-2131(s) 5.04 GLY-2324(b) 4.53 | LEU-2140(s) 5.02 GLY-236(b) 4.55 |
| C2 | THR-566(s) 4.14 GLY-840(b) 4.47 | ALA-501(s) 3.48 | LEU-374(s) 4.66 | ASP-637(s) 4.72 MET-921(s) 4.58 ASP-1754(s) 4.65 MET-2036(s) 4.57 ASP-2867(s) 4.67 MET-3141(s) | ALA-506(s) 3.57 | ALA-508(s) 3.58 | GLN-1000(s) 3.82 | ALA-516(s) 3.58 | GLN-1015(s) 3.85 | GLN-997(s) 3.82 | GLN-994(s) 3.92 |   | GLY-904(b) 4.28 THR-3245(s) 3.89 GLY-6105(b) 4.19 GLY-8704(b) 4.42 | ALA-499(s) 3.59 | ALA-508(s) 3.60 | ALA-520(s) 3.58 |
|   | THR-566(s) 4.65 TYR-876(s) 4.03 | GLY-505(b) 4.48 | LEU-374(s) 4.02 | SER-83(b) 4.44 SER-85(b) 3.29 MET-921(s) 5.53 SER-1201(b) 4.49 SER-1203(s) |   |   |   |   |   |   |   | LEU-423(s) 4.72 | THR-631(s) 4.38 TYR-936(s) 4.28 THR-3245(s) 4.41 TYR-3536(s) 3.79 | | GLY-512(b) 4.49 | GLY-524(b) 4.46 |

HBO | VDW

**Figure 11: Spreadsheet for analyzing the ligand binding site in the training structures** (See attached spreadsheet Analysis.xls for the full spreadsheet) : This spreadsheet lists the amino acids side chains or backbones that interacted with the different sites in the ligand. The distance in Angstroms between the sites and the protein were also tabulated so that they could be studied. From this spreadsheet the most dominant sites of the ligand **that interacted with the protein were identified. This was done by observing the interactions for the different proteins having FMN bound to it in the spreadsheet and tabulating the interacting atoms that were prevalent across all the proteins**

For FMN they were found to be:

1. C2
2. C6
3. O2P
4. O3P

However, when these four sites were considered a lot of false positives were identified for proteins that had GTP/GDP bound to them (Table 3). This was due to the fact that the phosphate group was decreasing the specificity of the algorithm. Hence the phosphate group binding site of FMN was
not considered as one of the atoms forming the edge of the tetrahedron that were interacting with the amino acid atoms of the sidechain or backbone of the protein. Thus the atoms of FMN that had predominant interactions with the protein molecules were now:

1. C2
2. C6
3. C5a
4. O2

These atoms are highlighted in Fig. 12

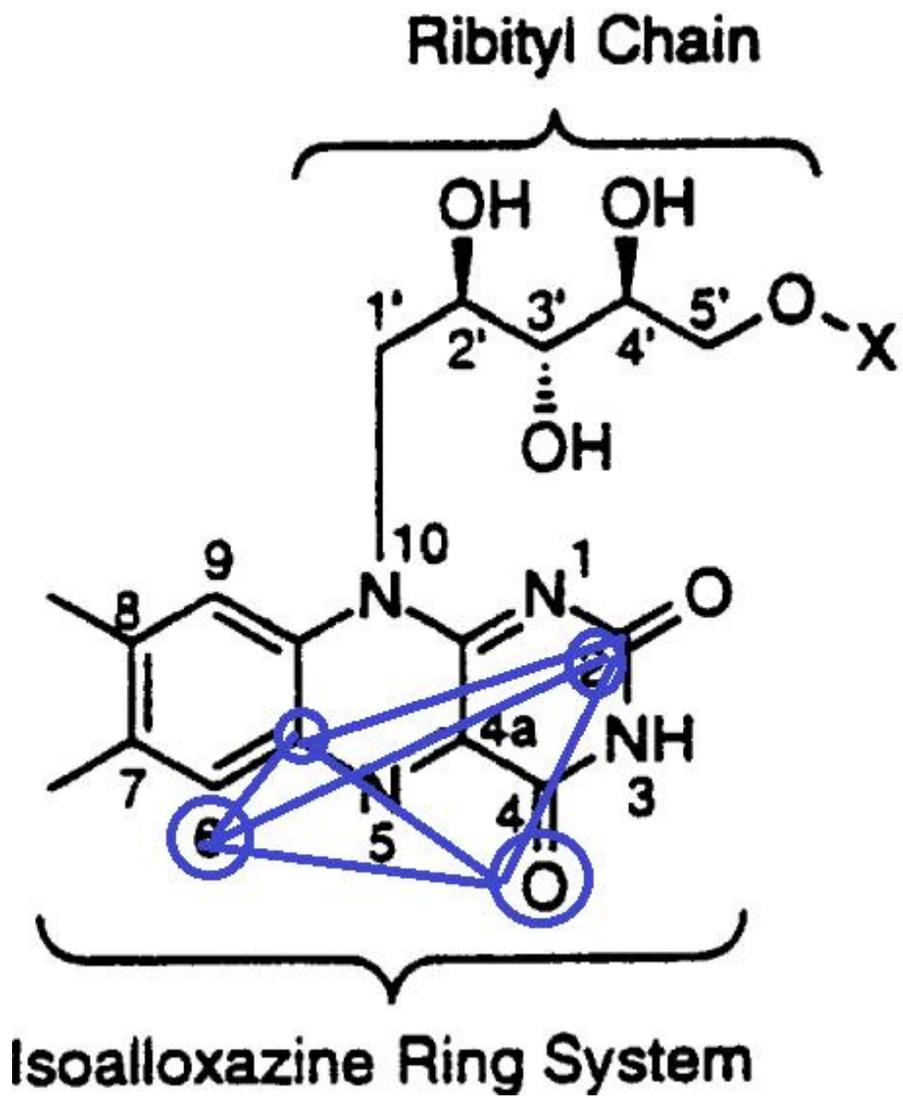

**Figure 12: Tetrahedral Motif in FMN**

So the tetrahedral motif was identified to be in the isoalloxazine ring of FMN. These would serve as the 4 nodes of the tetrahedral motif. The screening algorithm needs one of the nodes to be assigned as the root. This can be any of the four nodes. What this root node does is act as a starting point from which all the intermolecular interactions are found out. Our screening algorithm is written in Fortran77 and 90, and requires the parameters of the 3DSM, as input (i.e., the identities of the amino acid residues making up the 3D Screening Motif, the lengths of its 6 sides, and the nature of their interaction with the ligand, i.e., via backbone or via side chain, a total of at least

14 parameters). Candidate 3D Screening Motif 'sides' are sequestered by the algorithm from the test protein structure 3D structure based on the input parameters as well as the 'connectivity' (i.e., whether each node is connected to the same root, as well as to each other). In ascertaining

connectivity, groups of centroids called' clusters' are first selected, from which groups called 'trees'

are further selected. An 'error margin', ,, $\pm 1.4$ Å, is added to the lengths of the sides of the 3D SM to incorporate a fuzzy element into the screening process.[11]. If we do not give an error margin then we are making the dimensions of the tetrahedron very specific and not accounting for any small difference of the position of the atoms making up the nodes of the tetrahedral model. This would lead to a lot of false negatives. The error margin was chosen as 1.4 Å, to get a good rate of specificity and sensitivity. (Table 4, 5, 6)

O2 was chosen as the root node, C2 as n1, C6 as n2 and C5a as n3.

### Identifying the amino acids in the binding site that bound the ligand atom

Once the tertrahedral motif was identified it was necessary to find out what were the amino acids that were present in the sidechain or backbone of the protein molecule that were interacting with the ligand molecule. Again this was found by analyzing the spreadsheet that was prepared. In many cases there was more than one amino acid that was found to be common in the proteins for each node. The algorithm used allows for this. If there was more than one amino acid predominant they were all taken into consideration. The table below shows the predominant amino acids interacting with each node in the ligand.

| Node | Amino Acids |
|------|-------------|
| C2   | ALA(sidechain), GLN(sidechain) |
| C6   | TYR (sidechain), THR (sidechain) |
| C5a  | THR(backbone), GLY (backbone) |
| O2   | ASN(sidechain), LYS(sidechain) |

Table 2: Predominant amino acids interaction with each node in the ligand

This represents the tetrahedral motif model for the ligand binding site in the double centroid reduced representation.

The data gathered were sufficient to proceed with the screening step in which the algorithm would be trained using a few test structures.

## Verifying the screening algorithm

The screening algorithm takes the tetrahedral motif that was constructed above to capture the conserved ligand binding site consensus. The four most dominant interacting atoms make up the four nodes of the tetrahedron. One of the nodes is arbirtraily chosen as the root R and the others as the three nodes n1, n2 and n3. The distances between the nodes give us the exact dimensions of the tetrahedral motif. Thus the six distances Rn1, Rn2, Rn3, n1n2, n2n3 and n1n3 are the quantitative parameters that construct the fixed tetrahedral screening motif (Fig. 15).

Once we have the tetrahedral motif we try and identify where the motif can be found in the structure of the protein to be screened. If the tetrahedral motif does not exist then obviously the protein is eliminated and is considered as a negative for the screen. So the first step is to look for the amino acids in the sidechain or backbone that interact with the four nodes (Table 2). As in our case there can be more than one dominant amino acid interactions for each node. The screening algorithm allows for this. So in the first step we try to eliminate all the sites which do not have the dominant amino acids interacting with the four nodes (Fig. 16). Now these amino acids can be coming from the sidechain or backbone of the protein. Since they can come from either one for each amino acid we further eliminate some sites because they are not coming from the backbone of the protein or maybe the sidechain of the protein (Fig 17).

Now that the possible sites have been found in a qualitative manner we have actually identified the clusters in the protein structure which may have the tetrahedral motif we constructed from the training structures. However we do not really want all the clusters. The tetrahedral motif is now considered quantitatively to see if the six quantitative parameters discussed above match. Of course we do not consider the nodes and the roots as fixed dimensionless points. Variations occur among proteins and if we just had a tetrahedral of one fixed size then we will not get any positives. To account for this we create a small sphere of +- 1.4 angstroms around the root and the three nodes (fuzzy factor). We account for all lengths that satisfy this. We calculate all the distances among the amino acids and construct the trees from these clusters having the allowable distances (Fig 18).

The last step would be to match these edge nodes distances to the set of allowable distances we found above and see if the tetrahedral motif exists in the protein (Fig 19, 20).

Fortran programs achieve all of the above steps. If there is no match found then the program generates a blank output for that step.

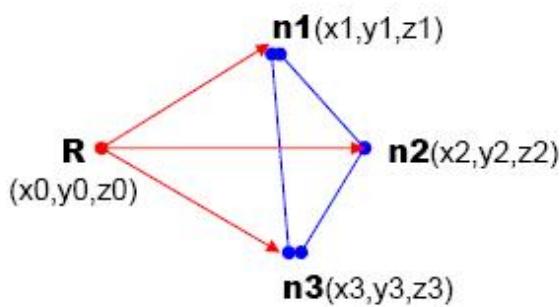

**General 3D Search Motif Structure:**

**Screening Algorithm Parameters:**

I. Qualitative parameters: at least 4; amino acid identities of:

   **R** == root

   **n1, n2, n3** == nodes

II. Quantitative parameters: at least 6; lengths of:

   **Rn1, Rn2, Rn3** == branches

   **n1n2, n2n3, n3n1** == node-edges

III. Fuzzy factor added to branch and node-edge lengths: ± ε (~1.40 Å)

**Figure 13: Creating an irregular tetrahedral motif to be used for screening [11]**

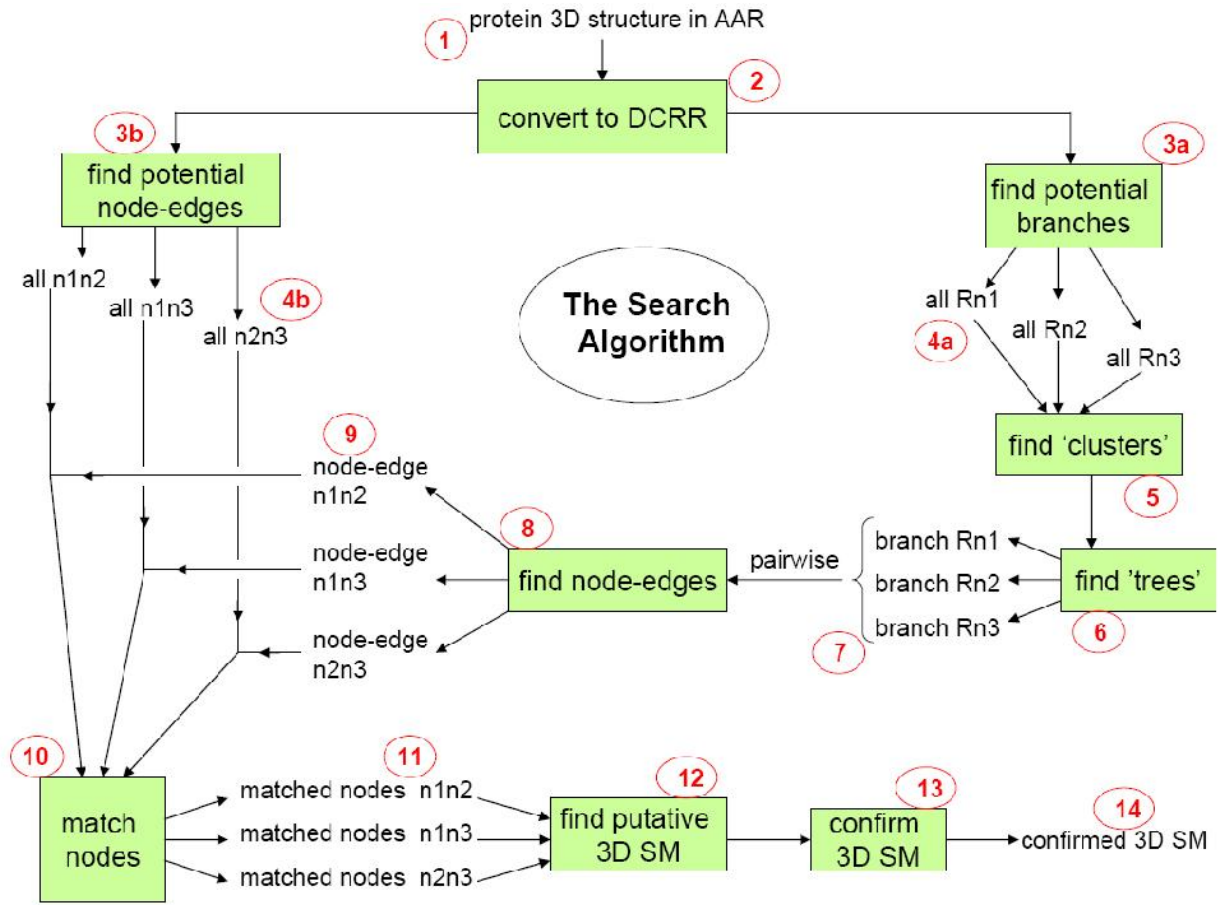

**Figure 14: Screening algorithm [11]**

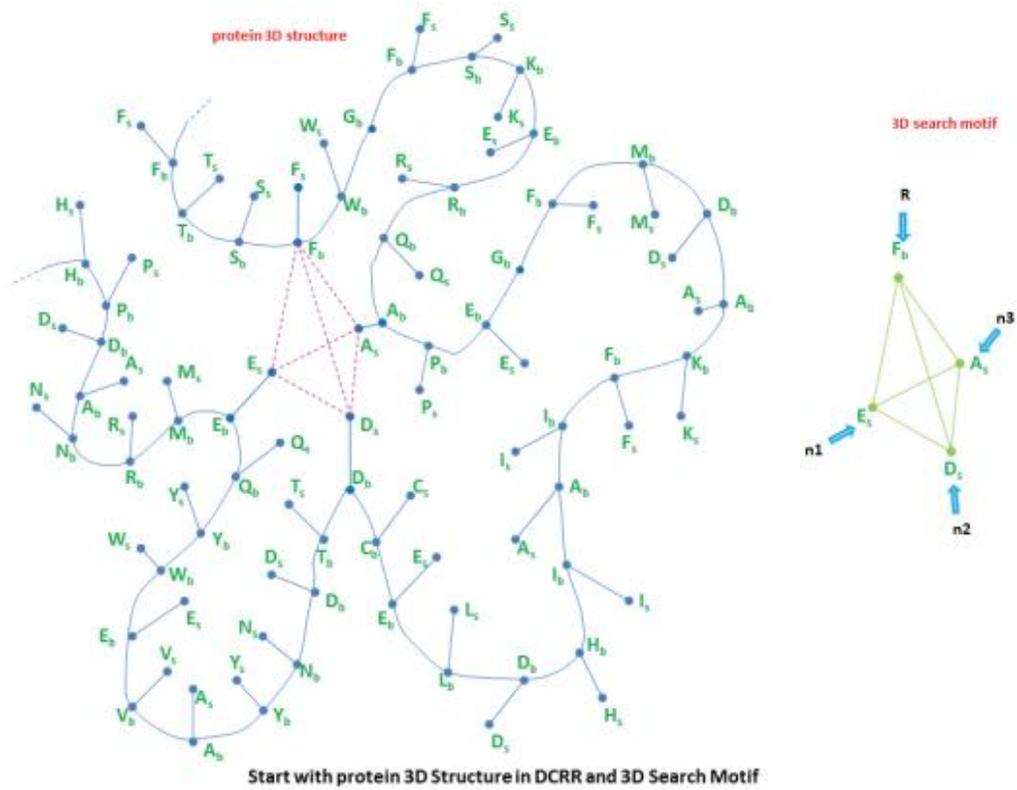

**Figure 15: Screening Algorithm Preparation** [11]: The entire protein structure of the protein to be tested in DCRR and the identified tetrahedral motif is taken.

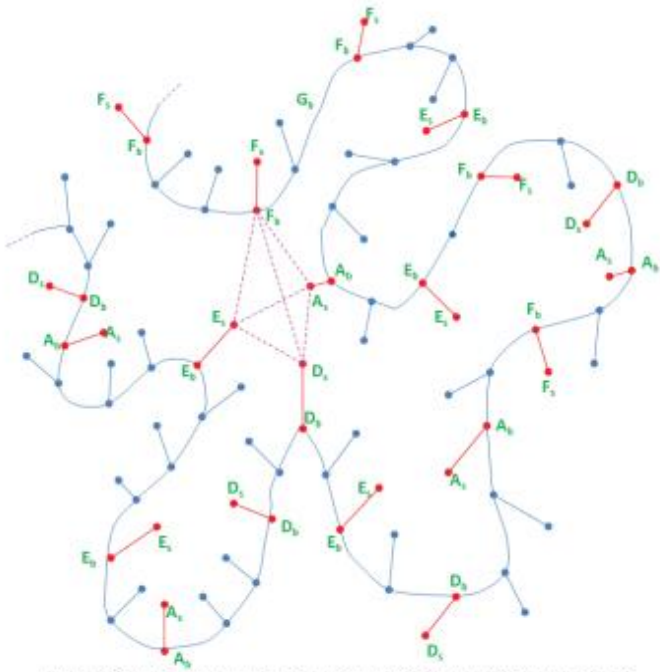

Step #1: Sequester amino acid residues in protein included in 3D Search Motif

**Figure 16: Screening Algorithm** Step1 [11] : Only the amino acids that were identified in the tetrahedral motif are considered in the structure of the protein to be tested

Step #2: Select backbone or side chain centroids according to 3D Search Motif

**Figure 17: Screening Algorth Step 2** [11]: The backbones and side chain centroids of the test protein are identified

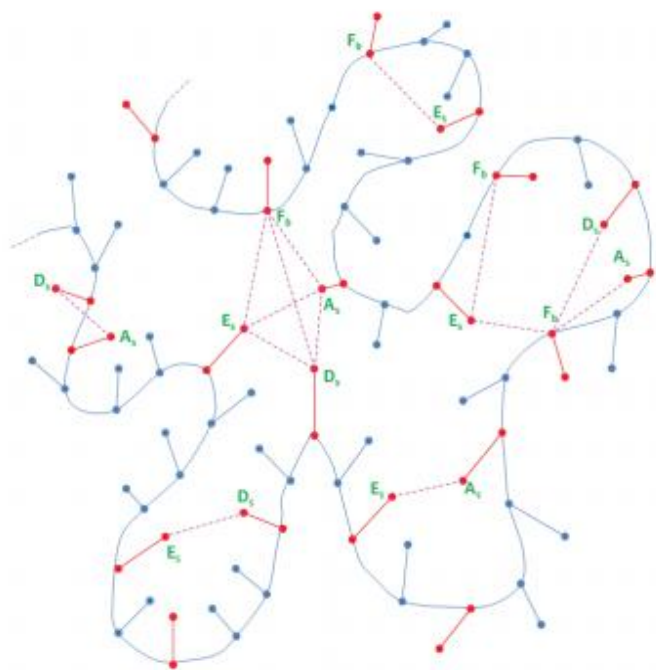

**Figure 18: Screening Algorithm Step3** [11]: The distances between the interacting atoms are calculated and only those within the limits of the tetrahedral model are selected

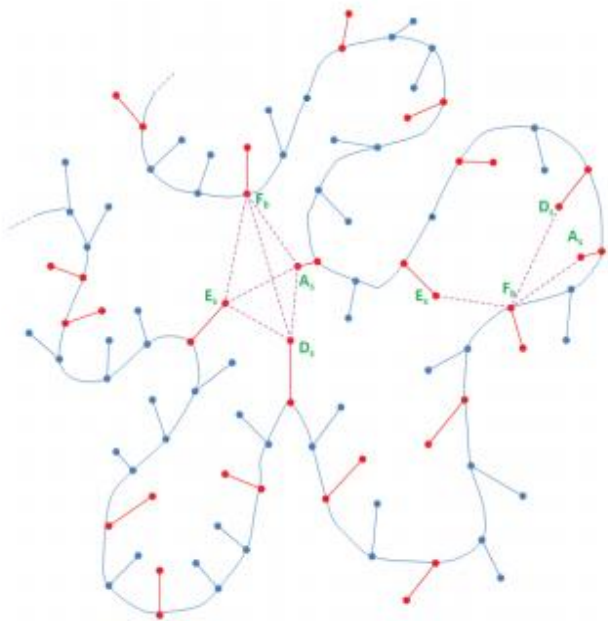

**Figure 19: Screening Algorithm Step 4** [11]: The roots and the nodes are chosen from the protein structures

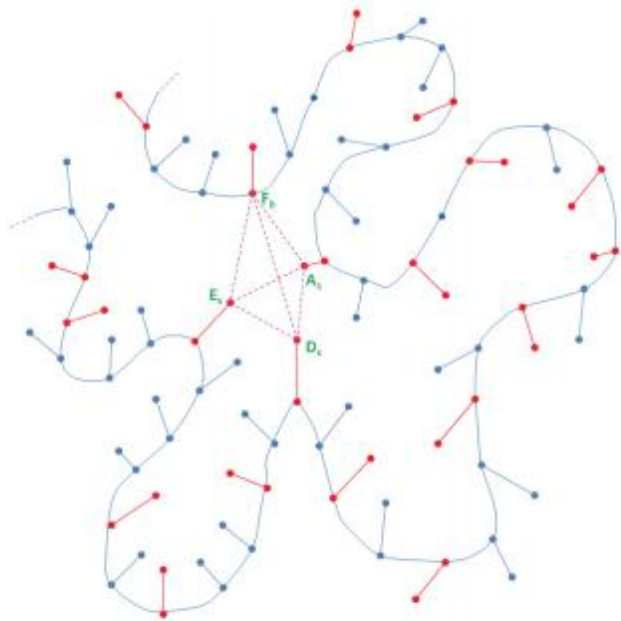

**Figure 20: Screening Algorthm Step 5** [11]: The node-edges that have lengths that are within limits of the 3D search motif are selected

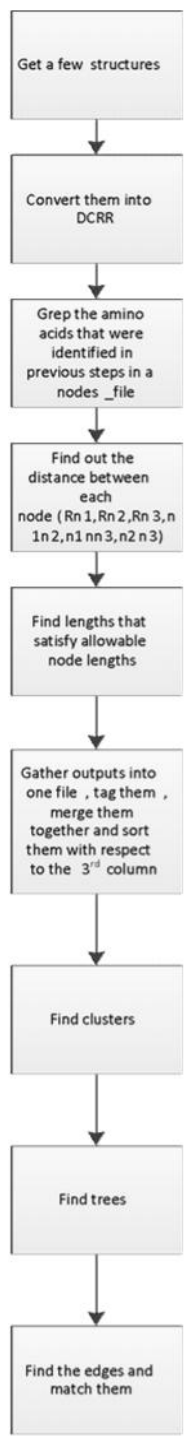

**Figure 21: Screening algorithm flow:** This flowchart describes the steps in the screening algorithm only. The different steps that the algorithm follows to screen a protein is described.

### Get a few structures

The Sixteen training structures were used to train the algorithm. The parameters derived from the training structures were fed into the algorithm. These parameters are used to create the tetrahedral motif. These include the qualitative parameters i.e. the ligand atoms that form the four nodes of the tetrahedral (the root and the three nodes) and also the quantitative parameters that set the dimensions of the tetrahedral motif.

Once the algorithm is trained it was tested against structures with known ligand binding sites to test for true positives, true negatives, false positives and false negatives. (Table 3, 4, 5, 6)

### Convert them into DCRR

These structures were then converted to DCRR structures from AAR structures using Vrunda Seth's algorithm.

### Prepare nodes_file

Once the DCRR structures for the proteins were ready a simple grep was performed to get only those amino acids which were identified in the pre-screening steps above. These files made up the input nodes files for the next steps.

### Find out the distance between each node

The distance between each node was found using the fortran program nrst_nghbr.f. The distances were separated into different files for the distances Rn1, Rn2, Rn3, n1n2, n2n3, n1n3.

### Find lengths that satisfy allowable node lengths

Using the median length found in the previous step for each group the lengths that satisfy allowable node lengths was computed using the six fortran programs dist_e1e2.f, dist_e1e3.f, dist_e2e3.f, dist_Re1.f, dist_Re2.f, dist_Re3.f. The allowable node lengths account for the fuzzy factor as described above

### Output gathered into one file, tagged merged together and sorted

The outputs from the above programs were first gathered into one file for each node length if there were multiple outputs. Then the files were tagged to indicate what node length the output was from (Rn1, Rn2, Rn1, n1n2, n2n3 or n1n2) and merged together into one file. This file was then sorted to do a group by all the node lengths.

### Find clusters

Using the output file from the previous step the clusters were found using the fortran program find_clstrs_3_nodes.f90.

### Find trees

Using the output of the clusters file above the trees were generated using the fortran program find_trees_3_nodes.f90

### Find the edges

Once the output above was generated all the branches (Rn1, Rn2, Rn3) were gathered together using a simple grep. Then the find_edgenode.f fortran program was used to find the edges (n1n2, n1n3, n2n3). These edges were matched with the branches to create the tetrahedral in the protein molecule. Thus the tetrahedral motif was confirmed for FMN.

| FAD bound protein | Tested Positive? | FMN bound protein | Tested Positive? | GDP/GTP bound protein | Tested Positive? |
|---|---|---|---|---|---|
| 3MBG | No | 1D3G | Yes | 3BRW | No |
| 3QJ4 | No | 1B1C | Yes | 1MH1 | No |
| 2CFY | No | 1NRG | Yes | 1JAH | No |
| 2QTL | No | 2BXV | Yes | 1JAI | No |
| 3QFS | No | 2RDU | Yes | 3ZYS | No |
| 3QFT | No | 2RDW | Yes | 3KUC | No |
| 2VZ2 | Yes | 2PRH | Yes | 1GNP | No |
| 2QTZ | Yes | 2RDT | Yes | 1GNQ | No |
| 3O55 | No | 3HR4 | Yes | 1GNR | No |
| 2UXX | No | 3KVJ | Yes | 1KAO | No |
| 2BK3 | No | 2NZL | Yes | 3CF6 | No |
| 3NKS | No | 2FPY | Yes | 3KUB | No |
| 3AHQ | No | 1QZU | Yes | 1FB1 | Yes |
| 3AHR | No | 3KVK | Yes | 3NC1 | No |
| 3QFC | No | 3KVM | Yes | 3NBY | No |
| 3QE2 | No | 2WKP | Yes | 3NBZ | No |
| 3QFB | No | 1HE4 | Yes | 3NCO | No |
| 3QFA | No | 2WV8 | Yes | 3Q7Q | No |
| 2WBI | No | 3QFC | Yes | 3Q72 | No |
| 2BYB | No | 3QE2 | Yes | 3Q7P | No |
| 3QFR | No | 3QFR | Yes | 1FBX | No |
| 2VRM | No | 3G0U | Yes | 2Q3F | No |
| 2VRL | No | 1D3H | Yes | 2CLS | No |
| 1QR2 | No | 3F1Q | Yes | 2J1L | No |
| 1GOS | No | 3FJ6 | Yes | 2EKI | No |
| 2CKJ | No | 3FJL | Yes | 2OCB | No |
| 2BXR | Yes | 3G0X | Yes | 3GFT | No |
| 2BXS | Yes | 2PRL | Yes | 2G6B | No |
| 3NHF | No | 2PRM | Yes | 3LLU | No |
| 3NHJ | No | 2W0U | Yes | 1KHE | No |
| 3NHK | No | 2Q9U | Yes | 1KHB | No |
| 3NHL | No | 2BOM | Yes | 1GUA | No |
| 3NHP | No | 1P4M | Yes | 2GMV | No |
| 3NHR | No | 3HY8 | Yes | 1WQ1 | No |
| 3NHS | No | 3KVL | Yes | 1TBG | No |
| 3NHU | No | 3FJO | Yes | 1AM4 | No |
| 3NHW | No | 3U2O | Yes | 2BC9 | No |
| 3NHY | No | 1Q9S | Yes | 1ZW6 | No |
| 3ZYX | No | 2WKQ | Yes | 2A5D | No |
| 1UMK | Yes | 2WKR | Yes | 2EW1 | No |
| 2H94 | No | 2FPT | Yes | 1A2B | Yes |
| 2VIG | No | 2FPV | Yes | 2FG5 | No |
| 2JIF | No | 2FPY | Yes | 3T5G | Yes |
| 2J3N | No | 2FQI | Yes | 1LF0 | No |
| 3PO7 | No | | | 1M51 | No |
| 1ZX1 | No | | | 1NHX | No |
| 1OJA | No | | | 1XTS | No |
| 2AAQ | No | | | 3EA5 | Yes |
| 2UXW | No | | | 1KHF | No |
| 3DJG | No | | | 1KHG | No |
| 3DK4 | No | | | 1QRA | No |
| 3DK9 | No | | | 3RAP | No |
| 2QR2 | No | | | 2GIL | No |
| 2F5Z | No | | | 2RAP | No |
| 2IW5 | No | | | 2B8W | No |
| 3DK8 | No | | | 2B92 | No |
| 3DJJ | No | | | 2D4H | No |
| 2Z5X | No | | | 2G3Y | No |
| 2Z5Y | No | | | 3CBQ | No |
| 3GOU | No | | | 1RRP | No |
| 3F1Q | No | | | 1DOA | No |
| 3FJ6 | No | | | 3P32 | Yes |
| 3FJL | No | | | 1K5D | No |
| 3G0X | No | | | 1K5G | No |
| 3LF5 | No | | | 3TW4 | Yes |
| 3H8Q | No | | | 2J5X | No |
| 3DU8 | No | | | 2DPX | No |
| 1T9G | Yes | | | 2GJS | No |
| 1S1Q | No | | | 1X3S | No |
| 2ZZC | No | | | 2GAO | No |
| 2ZZO | No | | | 2HT6 | No |
| 2ZZB | No | | | 1RA7 | No |
| 1SIR | No | | | 2OHF | No |
| 1OJD | No | | | 3T5I | No |
| 1OJ9 | No | | | 2A5F | No |

Table 3: When the algorithm was tested with the nodes as C2, C6, O2P and O3P the algorithm incorrectly identified a lot of structures having GTP/GDP bound to it as a false positive.

| FAD bound protein | Tested Positive? | FMN bound protein | Tested Positive? | GDP/GTP bound protein | Tested Positive? |
|---|---|---|---|---|---|
| 3MBG | No | 1D3G | Yes | 3BRW | No |
| 3QJ4 | No | 1B1C | Yes | 1MH1 | No |
| 2CFY | No | 1NRG | Yes | 1JAH | No |
| 2QTL | No | 2BXV | Yes | 1JAI | No |
| 3QFS | No | 2RDU | Yes | 3ZYS | No |
| 3QFT | No | 2RDW | Yes | 3KUC | No |
| 2VZ2 | No | 2PRH | Yes | 1GNP | No |
| 2QTZ | No | 2RDT | Yes | 1GNQ | No |
| 3O55 | No | 3HR4 | Yes | 1GNR | No |
| 2UXX | No | 3KVJ | Yes | 1KAO | No |
| 2BK3 | No | 2NZL | Yes | 3CF6 | No |
| 3NKS | No | 2FPY | Yes | 3KUB | No |
| 3AHQ | No | 1QZU | Yes | 1FB1 | No |
| 3AHR | No | 3KVK | Yes | 3NC1 | No |
| 3QFC | No | 3KVM | Yes | 3NBY | No |
| 3QE2 | No | 2WKP | Yes | 3NBZ | No |
| 3QFB | No | 1HE4 | Yes | 3NCO | No |
| 3QFA | No | 2WV8 | Yes | 3Q7Q | No |
| 2WBI | No | 3QFC | Yes | 3Q72 | No |
| 2BYB | No | 3QE2 | Yes | 3Q7P | No |
| 3QFR | No | 3QFR | Yes | 1FBX | No |
| 2VRM | No | 3G0U | Yes | 2Q3F | No |
| 2VRL | No | 1D3H | Yes | 2CLS | No |
| 1QR2 | No | 3F1Q | Yes | 2J1L | No |
| 1GOS | No | 3FJ6 | Yes | 2EKI | No |
| 2CKJ | No | 3FJL | Yes | 2OCB | No |
| 2BXR | Yes | 3G0X | Yes | 3GFT | No |
| 2BXS | Yes | 2PRL | Yes | 2G6B | No |
| 3NHF | No | 2PRM | Yes | 3LLU | No |
| 3NHJ | No | 2W0U | Yes | 1KHE | No |
| 3NHK | No | 2Q9U | Yes | 1KHB | No |
| 3NHL | No | 2BOM | Yes | 1GUA | No |
| 3NHP | No | 1P4M | Yes | 2GMV | No |
| 3NHR | No | 3HY8 | Yes | 1WQ1 | No |
| 3NHS | No | 3KVL | Yes | 1TBG | No |
| 3NHU | No | 3FJO | Yes | 1AM4 | No |
| 3NHW | No | 3U2O | Yes | 2BC9 | No |
| 3NHY | No | 1Q9S | Yes | 1ZW6 | No |
| 3ZYX | No | 2WKQ | Yes | 2A5D | No |
| 1UMK | No | 2WKR | Yes | 2EW1 | No |
| 2H94 | No | 2FPT | Yes | 1A2B | No |
| 2VIG | Yes | 2FPV | Yes | 2FG5 | No |
| 2JIF | Yes | 2FPY | Yes | 3T5G | No |
| 2J3N | Yes | 2FQI | Yes | 1LF0 | No |
| 3PO7 | No | | | 1M51 | No |
| 1ZX1 | Yes | | | 1NHX | No |
| 1OJA | Yes | | | 1XTS | No |
| 2AAQ | No | | | 3EA5 | No |
| 2UXW | No | | | 1KHF | No |
| 3DJG | No | | | 1KHG | No |
| 3DK4 | No | | | 1QRA | No |
| 3DK9 | No | | | 3RAP | No |
| 2QR2 | No | | | 2GIL | No |
| 2F5Z | No | | | 2RAP | No |
| 2IW5 | No | | | 2B8W | No |
| 3DK8 | No | | | 2B92 | No |
| 3DJJ | No | | | 2D4H | No |
| 2Z5X | No | | | 2G3Y | No |
| 2Z5Y | No | | | 3CBQ | No |
| 3GOU | No | | | 1RRP | No |
| 3F1Q | No | | | 1DOA | No |
| 3FJ6 | No | | | 3P32 | No |
| 3FJL | No | | | 1K5D | No |
| 3G0X | No | | | 1K5G | No |
| 3LF5 | No | | | 3TW4 | No |
| 3H8Q | No | | | 2J5X | No |
| 3DU8 | No | | | 2DPX | No |
| 1T9G | Yes | | | 2GJS | No |
| 1S1Q | No | | | 1X3S | No |
| 2ZZC | Yes | | | 2GAO | No |
| 2ZZO | Yes | | | 2HT6 | No |
| 2ZZB | Yes | | | 1RA7 | No |
| 1SIR | No | | | 2OHF | No |
| 1OJD | No | | | 3T5I | No |
| 1OJ9 | No | | | 2A5F | No |

Table 4: When the algorithm was tested with the nodes as C2, C6, O2 andC5a and an error margin of 1.8 angstroms a lot of structures having FAD as a ligand and not FMN was picked up. Thus the false positive rate was high.

| FAD bound protein | Tested Positive? | FMN bound protein | Tested Positive? | GDP/GTP bound protein | Tested Positive? |
|---|---|---|---|---|---|
| 3MBG | No | 1D3G | Yes | 3BRW | No |
| 3QJ4 | No | 1B1C | Yes | 1MH1 | No |
| 2CFY | No | 1NRG | Yes | 1JAH | No |
| 2QTL | No | 2BXV | Yes | 1JAI | No |
| 3QFS | No | 2RDU | Yes | 3ZYS | No |
| 3QFT | No | 2RDW | Yes | 3KUC | No |
| 2VZ2 | No | 2PRH | Yes | 1GNP | No |
| 2QTZ | No | 2RDT | Yes | 1GNQ | No |
| 3O55 | No | 3HR4 | Yes | 1GNR | No |
| 2UXX | No | 3KVJ | Yes | 1KAO | No |
| 2BK3 | No | 2NZL | Yes | 3CF6 | No |
| 3NKS | No | 2FPY | Yes | 3KUB | No |
| 3AHQ | No | 1QZU | Yes | 1FB1 | No |
| 3AHR | No | 3KVK | Yes | 3NC1 | No |
| 3QFC | No | 3KVM | Yes | 3NBY | No |
| 3QE2 | No | 2WKP | Yes | 3NBZ | No |
| 3QFB | No | 1HE4 | Yes | 3NCO | No |
| 3QFA | No | 2WV8 | Yes | 3Q7Q | No |
| 2WBI | No | 3QFC | Yes | 3Q72 | No |
| 2BYB | No | 3QE2 | Yes | 3Q7P | No |
| 3QFR | No | 3QFR | Yes | 1FBX | No |
| 2VRM | No | 3G0U | Yes | 2Q3F | No |
| 2VRL | No | 1D3H | Yes | 2CLS | No |
| 1QR2 | No | 3F1Q | Yes | 2J1L | No |
| 1GOS | No | 3FJ6 | Yes | 2EKI | No |
| 2CKJ | No | 3FJL | Yes | 2OCB | No |
| 2BXR | Yes | 3G0X | Yes | 3GFT | No |
| 2BXS | Yes | 2PRL | Yes | 2G6B | No |
| 3NHF | No | 2PRM | Yes | 3LLU | No |
| 3NHJ | No | 2W0U | Yes | 1KHE | No |
| 3NHK | No | 2Q9U | Yes | 1KHB | No |
| 3NHL | No | 2BOM | Yes | 1GUA | No |
| 3NHP | No | 1P4M | Yes | 2GMV | No |
| 3NHR | No | 3HY8 | Yes | 1WQ1 | No |
| 3NHS | No | 3KVL | Yes | 1TBG | No |
| 3NHU | No | 3FJO | Yes | 1AM4 | No |
| 3NHW | No | 3U2O | Yes | 2BC9 | No |
| 3NHY | No | 1Q9S | No | 1ZW6 | No |
| 3ZYX | No | 2WKQ | Yes | 2A5D | No |
| 1UMK | No | 2WKR | Yes | 2EW1 | No |
| 2H94 | No | 2FPT | Yes | 1A2B | No |
| 2VIG | No | 2FPV | Yes | 2FG5 | No |
| 2JIF | No | 2FPY | Yes | 3T5G | No |
| 2J3N | No | 2FQI | Yes | 1LF0 | No |
| 3PO7 | No | | | 1M51 | No |
| 1ZX1 | No | | | 1NHX | No |
| 1OJA | No | | | 1XTS | No |
| 2AAQ | No | | | 3EA5 | No |
| 2UXW | No | | | 1KHF | No |
| 3DJG | No | | | 1KHG | No |
| 3DK4 | No | | | 1QRA | No |
| 3DK9 | No | | | 3RAP | No |
| 2QR2 | No | | | 2GIL | No |
| 2F5Z | No | | | 2RAP | No |
| 2IW5 | No | | | 2B8W | No |
| 3DK8 | No | | | 2B92 | No |
| 3DJJ | No | | | 2D4H | No |
| 2Z5X | No | | | 2G3Y | No |
| 2Z5Y | No | | | 3CBQ | No |
| 3GOU | No | | | 1RRP | No |
| 3F1Q | No | | | 1DOA | No |
| 3FJ6 | No | | | 3P32 | No |
| 3FJL | No | | | 1K5D | No |
| 3G0X | No | | | 1K5G | No |
| 3LF5 | No | | | 3TW4 | No |
| 3H8Q | No | | | 2J5X | No |
| 3DU8 | No | | | 2DPX | No |
| 1T9G | Yes | | | 2GJS | No |
| 1S1Q | No | | | 1X3S | No |
| 2ZZC | No | | | 2GAO | No |
| 2ZZO | No | | | 2HT6 | No |
| 2ZZB | No | | | 1RA7 | No |
| 1SIR | No | | | 2OHF | No |
| 1OJD | No | | | 3T5I | No |
| 1OJ9 | No | | | 2A5F | No |

Table 5: When the algorithm was tested with the nodes as C2, C6, O2 andC5a and an error margin of 1.4 angstroms the selectivity and specificity of the algorithm was optimal.

| FAD bound protein | Tested Positive? | FMN bound protein | Tested Positive? | GDP/GTP bound protein | Tested Positive? |
|---|---|---|---|---|---|
| 3MBG | No | 1D3G | Yes | 3BRW | No |
| 3QJ4 | No | 1B1C | Yes | 1MH1 | No |
| 2CFY | No | 1NRG | Yes | 1JAH | No |
| 2QTL | No | 2BXV | Yes | 1JAI | No |
| 3QFS | No | 2RDU | Yes | 3ZYS | No |
| 3QFT | No | 2RDW | Yes | 3KUC | No |
| 2VZ2 | No | 2PRH | Yes | 1GNP | No |
| 2QTZ | No | 2RDT | Yes | 1GNQ | No |
| 3O55 | No | 3HR4 | No | 1GNR | No |
| 2UXX | No | 3KVJ | No | 1KAO | No |
| 2BK3 | No | 2NZL | Yes | 3CF6 | No |
| 3NKS | No | 2FPY | Yes | 3KUB | No |
| 3AHQ | No | 1QZU | Yes | 1FB1 | No |
| 3AHR | No | 3KVK | No | 3NC1 | No |
| 3QFC | No | 3KVM | No | 3NBY | No |
| 3QE2 | No | 2WKP | Yes | 3NBZ | No |
| 3QFB | No | 1HE4 | Yes | 3NCO | No |
| 3QFA | No | 2WV8 | Yes | 3Q7Q | No |
| 2WBI | No | 3QFC | Yes | 3Q72 | No |
| 2BYB | No | 3QE2 | Yes | 3Q7P | No |
| 3QFR | No | 3QFR | Yes | 1FBX | No |
| 2VRM | No | 3G0U | Yes | 2Q3F | No |
| 2VRL | No | 1D3H | Yes | 2CLS | No |
| 1QR2 | No | 3F1Q | Yes | 2J1L | No |
| 1GOS | No | 3FJ6 | Yes | 2EKI | No |
| 2CKJ | No | 3FJL | Yes | 2OCB | No |
| 2BXR | No | 3G0X | Yes | 3GFT | No |
| 2BXS | No | 2PRL | Yes | 2G6B | No |
| 3NHF | No | 2PRM | Yes | 3LLU | No |
| 3NHJ | No | 2W0U | Yes | 1KHE | No |
| 3NHK | No | 2Q9U | Yes | 1KHB | No |
| 3NHL | No | 2BOM | Yes | 1GUA | No |
| 3NHP | No | 1P4M | No | 2GMV | No |
| 3NHR | No | 3HY8 | Yes | 1WQ1 | No |
| 3NHS | No | 3KVL | Yes | 1TBG | No |
| 3NHU | No | 3FJO | Yes | 1AM4 | No |
| 3NHW | No | 3U2O | Yes | 2BC9 | No |
| 3NHY | No | 1Q9S | No | 1ZW6 | No |
| 3ZYX | No | 2WKQ | Yes | 2A5D | No |
| 1UMK | No | 2WKR | Yes | 2EW1 | No |
| 2H94 | No | 2FPT | Yes | 1A2B | No |
| 2VIG | No | 2FPV | Yes | 2FG5 | No |
| 2JIF | No | 2FPY | Yes | 3T5G | No |
| 2J3N | No | 2FQI | Yes | 1LF0 | No |
| 3PO7 | No | | | 1M51 | No |
| 1ZX1 | No | | | 1NHX | No |
| 1OJA | No | | | 1XTS | No |
| 2AAQ | No | | | 3EA5 | No |
| 2UXW | No | | | 1KHF | No |
| 3DJG | No | | | 1KHG | No |
| 3DK4 | No | | | 1QRA | No |
| 3DK9 | No | | | 3RAP | No |
| 2QR2 | No | | | 2GIL | No |
| 2F5Z | No | | | 2RAP | No |
| 2IW5 | No | | | 2B8W | No |
| 3DK8 | No | | | 2B92 | No |
| 3DJJ | No | | | 2D4H | No |
| 2Z5X | No | | | 2G3Y | No |
| 2Z5Y | No | | | 3CBQ | No |
| 3GOU | No | | | 1RRP | No |
| 3F1Q | No | | | 1DOA | No |
| 3FJ6 | No | | | 3P32 | No |
| 3FJL | No | | | 1K5D | No |
| 3G0X | No | | | 1K5G | No |
| 3LF5 | No | | | 3TW4 | No |
| 3H8Q | No | | | 2J5X | No |
| 3DU8 | No | | | 2DPX | No |
| 1T9G | Yes | | | 2GJS | No |
| 1S1Q | No | | | 1X3S | No |
| 2ZZC | No | | | 2GAO | No |
| 2ZZO | No | | | 2HT6 | No |
| 2ZZB | No | | | 1RA7 | No |
| 1SIR | No | | | 2OHF | No |
| 1OJD | No | | | 3T5I | No |
| 1OJ9 | No | | | 2A5F | No |

Table 6: When the algorithm was tested with the nodes as C2, C6, O2 andC5a and an error margin of 1.0 angstroms the dimensions of the tetrahedral structure became to specific and could not pick up a lot of proteins that actually had FMN bound to it. A low error margin increased the rate of false negatives.

## Screening unannotated structures of PDB

After the algorithm was trained a pipeline script was prepared for the steps above so that files from PDB that are unannotated could be quickly processed to see if the tertrahedral motif was present in them. These files were found by searching PDB for structures with an unknown function.

If the proteins did not survive the steps then a blank output was generated and hence that protein eliminated.

These are the proteins that survived the screening from 1372 proteins (5.17%) in PDB.

1. 1ovq – Hypothetical protein in E.Coli that could be a nuclease resolving Holliday junction intermediates in genetic recombination by being redox active coordination complexes and hence may have FMN binding motif.

2. 1pc2 - Human mitochondrial protein Fis1 having a high likelihood of having a FMN binding motif

3. 1q53 - Hypothetical protein At3g17210 from A. thaliana with unknown function.

4. 1rfl – E.Coli MnmE protein which is equivalent to proteins in eukaryotes crucial for mitochondrial respiration. Its homologues have FMN binding motifs.

5. 1tr4 - Human gankyrin which is known to complex with coenzyme FMN.

6. 1u3n – A prokaryotic superoxide dismutase paralog lacking two Cu ligands. Its homologues are known to have FMN and NADPH ligand binding motifs.

7. 1wix - Mouse Hook homolog1. Homologues have a FMN binding site.

8. 1xpn - PA1324 protein in Pseudomonas aeruginosa having unknown function

9. 1ydu - At5g01610 protein in Arabidopsis thaliana having unknown function.

10. 1yyc - putative late embryogenesis abundant protein with unknown function

11. 2asy - E. coli protein YdhR which likely belongs to a recently identified group of mono oxygenase proteins that have a FMN binding site.

12. 2do8 - UPF0301 protein HD_1794 of Haemophilus ducreyi having an unknown function

13. 2e63 - KIAA1787 protein of Drosophila having an unknown function

14. 2ec4 - FAF1 in humans that probably plays a role as an apoptotic signaling regulator and may have a

FMN binding site

15. 2fyw - Hypothetical protein in Streptococcus pneumonia having unknown function

16. 1iyg – Hypothetical protein in mouse having unknown function

17. 1j7h – Hypothetical protein HI0719 having unknown function

18. 1nxi - Hypothetical protein VC0424 from Vibrio cholera that is thought to behave like a ferrodoxin. A FMN binding site is very likely.

19. 1sgo - Protein C14orf129 gene product in humans having unknown function

20. 2g0i - Hypothetical protein SMU 848 in Streptococcus mutans with unknown function.

21. 2fne -  Multiple PDZ domain protein in humans having unknown function.

22. 2dcq -  Putative protein At4g01050 having unknown function.

23. 2dcp -  Hypothetical protein in A. thaliana having unknown function

24. 2daw – RWD domain containing protein 2 in humans that is thought to enhance the sumoylation of a number of proteins. A FMN binding site is very likely.

25. 2dax -  Protein C21orf6 in humans having unknown function

26. 2cq9 -  GLRX2 protein in humans having unknown function

27 2b3w -  Hypothetical protein ybiA in E.coli having unknown function

28. 1dm5 – Annexin XII E105k mutant homohexamer in Hydra vulgaris having unknown function

29. 1j31 -  Hypothetical protein PH0642 in E.coli having unknown function

30. 1jal -  YchF protein in Haemophilus influenza having unknown function

31. 1jri -  Sm like Archael Protein 1 in Methanothermobacter therautotrophicus having unknown function

32. 1kq4 - Hypothetical protein in Thermotoga maritime having unknown function

33. 1lj7 - Monomer hemoglobin component III in Glycera dibranchiate participating in oxygen storage/ transport. Likely to have a FMN binding motif.

34. 1lql - osmotical inducible protein C like family in Mycoplasma pneumonia having unknown function

35. 1m98 – Orange carotenoid protein in Arthospira maxima having unknown function

36. 1nf2 – Hypothetical protein in Thermotoga maritime having unknown function

37. 1nkq - Hypothetical protein in yeast having unknown function

38 1nmo - Hypothetical protein in ybgI having unknown function

39. 1nmp - Hypothetical protein in ybgI having unknown function

40. 1npy – Hypothetical protein in Haemophilus influenzae having unknown function

41. 1nr9 - Protein YCGM in E. coli with unknown function

42. 1nx8 - Carbapenem synthase of pectobacterium carotovorum having unknown function

43. 1nye – Osmotically inducible protein in E.coli having unknown function

44. 1o8c – Putative quinine oxidoreductase YHDH in E.coli. Very likely to have a FMN binding motif

45. 1oq1 -  Protein yesU in Bacillus subtilis having unknown function

46. 1oy1 – Putative sigma cross reacting protein 27A having unknown function

47. 1pt5. – Hypothetical protein yfdW in E.coli having unknown function.

48. 1pt7 - Hypothetical protein yfdW in E.coli having unknown function.

49. 1pt8 - Hypothetical protein yfdW in E.coli having unknown function.

50. 1qvv - YDR533c protein in yeast having unknown function

51. 1qv9 – F420 dependent methylenetetrahydromethanopterin dehrdrogenase in Methanopyrus kandleri. This behaves as an oxidoreductase and is very likely to have a FMN binding site.

52. 1rtw. – Putative transcriptional activator in Pyrococcus furiosus DSM 3638 having unknown function.

53. 1sg9 - hemK protein in Thermotoga maritima having unknown function

54. 1t0b - ThuA like protein in Geobacillus stearothermophilus having unknown function

55. 1t0t – APC35880 protein in Geobacillus stearothermophilus having unknown function

56. 1syr – thioredoxin of Plasmodium falcipurum that might act as an oxidoreductase and is likely to have a FMN binding site.

57. 1t2b – P450cin in Citrobacter braakii having unknown function.

58. 1t5r – LukS-PV in Staphylococcus phage having unknown function

59. 1tel – Ribulose biphosphate carboxylase in Chlorobium tepidum that is likely to have a FMN binding motif.

60. 1to0 - Hypothetical protein in Bacillus subtilis having unknown function

61. 1tt7 – YHFP protein in Bacillus subtilis having unknown function

62. 1twy– ABC transporter protein of Vibrio clolerae

63. 1u5w - Hypothetical protein in E.coli having unknown function

64. 1uc2 - Hypothetical protein in Pyrococcus horikoshii having unknown function

65. 1uf3 - Hypothetical protein in Thermus thermophilius having unknown function

66. 1v8p - Hypothetical protein in Pyrobaculum aerophilum having unknown function

67. 1v8o - Hypothetical protein in Pyrobaculum aerophilum having unknown function

68. 1vdh– Muconolactone isomerase like protein in Thermus thermophilus having unknown function

69. 1vhc – Putative KHG/KDPG aldolase in Haemophilus influenza having unknown function

70. 1vkd – Predicted glycosidase in Thermotoga maritima that is likely to have a FMN binding site.

71. 1wpb – Hypothetical protein in E.coli having unknown function

## DISCUSSION OF RESULTS:

Choosing the four nodes of the tetrahedral motif as amino acid sidechains or backbones that bind to C2,C6,C5a,O2 of FMN and having a error margin of 1.4 angstroms as discussed above led to a false positive rate of 4% and a false negative rate of 2.27%. Taking this into consideration of the 71 protein structures tested positive 68 might actually be true positives while of the 1301 structures that were rejected about 30 could have been rejected when they should have been picked up. However a lot of the structures tested were not from humans and since the training was done with structures from humans more work needs to be done to see how much the tetrahedral motif varies between different organisms. If the variation is large then the algorithm would have to be separately trained for each species.

Of the structures that were currently identified a lot of them were unknown proteins with unknown functions. However it is noteworthy to mention that the protein 1PC2 was picked up as a positive match. Now this protein is a mitochondrial fission protein. Though the ligand information is unknown normally mitochondrial fission proteins are known to have FMN involved [18]. Similarly 1wix is a mouse Hook homolog. Some of its homologues have been known to have FMN binding sites [19]. However in some cases the algorithm did not do well. Like 1rtw was identified as a positive hit. However it is putative transcriptional activator in Pyrococcus furiosus DSM 3638 and it is highly unlikely that there is a FMN binding site in it. 1m98 is an Orange carotenoid protein in Arthospira maxima and again it is highly unlikely that it has a FMN binding site even though the algorithm says so.

However the highly unlikely candidates are proteins that are found in species that are very diverse from humans and hence the algorithm is likely not trained suitably for a species that is not close to humans.

Since the algorithm suggests some likely proteins that possibly have FMN binding sites the next step would be to do wet lab experiments and see if the hypothesis is true. Strong candidates as the ones discussed above can act as a starting ground for future wet lab experiments and help in establishing the success of the algorithm. If the algorithm is found to have performed reasonably well it can be a ground breaking tool in identifying ligand binding sites for a particular ligand. Pharmacophore identification can then act as a tool for discovering novel molecules and drugs designed keeping the pharmacophore in mind.

# CONCLUSION:

The idea for this work came from the benefits that were presented before us from the DCRR model for protein representation. DCRR allowed us to simplify the protein structure information by reducing the overall number of data points by 70%. This was a step forward in identifying the pharmacophores which has remained a daunting task for years now. Using the DCRR a tetrahedral motif could be identified and that motif could be used to screen a large number of proteins quickly and efficiently. The DCRR also identifies the hydrogen bonding and Van der waals information in separate files and they are immensely useful to train the consensus ligand binding site for the pharmacophore identifier algorithm

The mathematical modeling of a ligand binding site can lead to new drug discoveries by promoting a way to screen normal proteins for those binding sites and thus revolutionize medicine. Using the algorithm developed and the pipeline script a molecule can be quickly screened to determine if the motif is present in it. This work serves as a very useful addon to the DCRR project and demonstrates the effectiveness of using DCRR to revolutionize the field of bioinformatics

# Challenges

This project was a very challenging one to me and gave me the opportunity to acquire new skill sets and programming knowledge. One of the big problems were the inconsistency of the PDB format which could sometimes throw the programs written for the algorithm off and not give outputs even though an output was expected. Just for this even if the output files were blank (meaning the protein was eliminated) I had to double check to see if there was a fault in the input files given and modify the programs accordingly. Several fortran programs had to be changed to recognize the new format before processing the data. More standardization in PDB would avoid this slowdown. One of the other challenges I faced was disk space and managing it efficiently to process several large files quickly. I had to use some servers from various places in order to accommodate this.

# FUTURE DIRECTIONS

This work was the first of its kind to utilize the DCRR files to find useful information. However a lot of the process had to be manually started off in various steps. If I had the chance to work on this project more

I would automate the whole process using newer technologies like Java/J2ee. This would not only speed the whole thing up but it would ensure that the output we get is more authentic. Several programs could be written to act as listeners and then check the format of the file and then modify the programs accordingly, recompile and then proceed with the steps. A web based interface could be designed where the user could input the required information and be notified of an output once the process is done.

Several other species could be considered while preparing the training site especially ones that are close to humans.

Also several other ligands like FAD, NAD etc could be considered and phramacophore templates found using the same methods. A library could then be maintained and made available to everyone for use in research.

neutron scattering and X-ray diffraction results, Published 4[th] January 2005, New J. Chem., 2005, 29 , 371–377 (www.rsc.org/njc),

http://chem.unipune.ernet.in/~tcg/members/1.pdf

*22. Reyes, V.M. "An Exact (Analytical) Algorithm for Structure-Based Protein Function Prediction via Detection of Specific Ligand 3D Binding Site Motifs" (2007a) (submitted & under revision)*

# Appendix

**Script 1:**

```
# Arkanjan Banerjee Step1
# October 2011, RIT

#!/bin/bash
grep "ASN" prot.dcrr | grep "scc" > nodes_file

grep "LYS" prot.dcrr | grep "scc" >> nodes_file grep "ALA"
prot.dcrr | grep "scc" >> nodes_file

grep "GLN" prot.dcrr | grep "scc" >> nodes_file grep "TYR"
prot.dcrr | grep "scc" >> nodes_file grep "THR" prot.dcrr | grep
"scc" >> nodes_file grep "THR" prot.dcrr | grep "bbc" >>
nodes_file grep "GLY" prot.dcrr | grep "bbc" >> nodes_file
cp   nodes_file    filea;        # creates the 'home' file cp
    nodes_file fileb;            # creates 'neighbor' file
./nrst_nghbr.x;                  # runs the nearest neighbor program mv fileo
dist_1
```

```
grep "ALA" dist_1 | grep "scc" >> Group1
grep "GLN" dist_1 | grep "scc" >> Group1

grep "TYR" dist_1 | grep "scc" >> Group2
grep "THR" dist_1 | grep "scc" >> Group2
grep "THR" dist_1 | grep "bbc" >> Group3
grep "GLY" dist_1 | grep "bbc" >> Group3
grep "ASN" dist_1 | grep "scc" >> Group4
grep "LYS" dist_1 | grep "scc" >> Group4

exit
```

**Script 2:**

# Arkanjan Banerjee Step2
# October 2011, RIT

```bash
#!/bin/bash

./dist_e1e3.x;
./dist_e1e2.x;
./dist_e2e3.x;
./dist_Re1.x;
./dist_Re2.x;
./dist_Re3.x;
cat Rn1 > Prot_all.Re1 ; cat Rn2 > Prot_all.Re2; cat Rn3 > Prot_all.Re3;
awk '{print $0" Re1"}' Prot_all.Re1 > Prot_all.Re1.t; awk '{print $0" Re2"}' Prot_all.Re2 > Prot_all.Re2.t; awk '{print $0" Re3"}' Prot_all.Re3 > Prot_all.Re3.t;

cat Prot_all.Re?.t > Prot_R_e1e2e3;

sort +2 -3 Prot_R_e1e2e3 > R_e1e2e3.s;

cp R_e1e2e3.s filei;     gfortran find_clstrs_3_nodes.f90;     ./a.out;     mv fileo R_e1e2e3.clstrs;     \rm filei    fc.f90 a.out;

cp R_e1e2e3.clstrs filei;     gfortran ft_3nodes_.f90;     ./a.out;     mv fileo
```

```
R_e1e2e3.trees;        \rm filei a.out;

sort +7 -8 R_e1e2e3.trees > R_e1e2e3.trees.s;

grep ' Re1' R_e1e2e3.trees.s > R_e1e2e3.trees.branch_Re1; grep ' Re2' R_e1e2e3.trees.s > R_e1e2e3.trees.branch_Re2; grep ' Re3' R_e1e2e3.trees.s > R_e1e2e3.trees.branch_Re3;

cp R_e1e2e3.trees.branch_Re1 filea; cp R_e1e2e3.trees.branch_Re2 fileb; ./fen.x; mv fileo edgenode_e1e2; \rm file[a,b];

cp R_e1e2e3.trees.branch_Re1 filea; cp R_e1e2e3.trees.branch_Re3 fileb; ./fen.x; mv fileo edgenode_e1e3; \rm file[a,b];

cp R_e1e2e3.trees.branch_Re2 filea; cp R_e1e2e3.trees.branch_Re3 fileb; ./fen.x; mv fileo edgenode_e2e3; \rm file[a,b];

exit
```